\documentclass[aps,reprint,twocolumn,superscriptaddress,citeautoscript,amsmath,amssymb,floatfix,longbibliography,prl]{revtex4-2}
\usepackage{graphicx}
\usepackage{times}
\usepackage{bm}
\usepackage{color}
\usepackage{gensymb}
\usepackage{dcolumn}
\usepackage{xfrac}
\usepackage{braket}
\usepackage[labelfont=bf]{caption}
\usepackage{soul}

\usepackage[resetlabels]{multibib}

\usepackage{xr}
\newcolumntype{d}[1]{D{.}{.}{#1}}

\begin{document}

\title{Nature of quantum criticality in the Ising ferromagnet TbV$_{\bm{6}}$Sn$_{\bm{6}}$}
\author{Tianxiong~Han}
	\affiliation{Ames National Laboratory, U.S. DOE, Iowa State University, Ames, Iowa 50011, USA}
	\affiliation{Department of Physics and Astronomy, Iowa State University, Ames, Iowa 50011, USA}
 \author{R. D. McKenzie}
	\affiliation{Ames National Laboratory, U.S. DOE, Iowa State University, Ames, Iowa 50011, USA}
	\affiliation{Department of Physics and Astronomy, Iowa State University, Ames, Iowa 50011, USA}
 \author{Joanna Blawat}
	\affiliation{National High Magnetic Field Laboratory, Los Alamos National Laboratory, Los Alamos, New Mexico 87545, USA}
 \author{Tyler J. Slade}
	\affiliation{Ames National Laboratory, U.S. DOE, Iowa State University, Ames, Iowa 50011, USA}
  \author{Bing Li}
  	\affiliation{Ames National Laboratory, U.S. DOE, Iowa State University, Ames, Iowa 50011, USA}
    \affiliation{Oak Ridge National Laboratory, Oak Ridge, TN, 37831, USA}
  \author{Y. Lee}
	\affiliation{Ames National Laboratory, U.S. DOE, Iowa State University, Ames, Iowa 50011, USA}
 \author{D. M. Pajerowski}
\affiliation{Oak Ridge National Laboratory, Oak Ridge, TN, 37831, USA}
 \author{John Singleton}
	\affiliation{National High Magnetic Field Laboratory, Los Alamos National Laboratory, Los Alamos, New Mexico 87545, USA}
 \author{Paul C. Canfield}
	\affiliation{Ames National Laboratory, U.S. DOE, Iowa State University, Ames, Iowa 50011, USA}
	\affiliation{Department of Physics and Astronomy, Iowa State University, Ames, Iowa 50011, USA}
  \author{Liqin Ke}
	\affiliation{Ames National Laboratory, U.S. DOE, Iowa State University, Ames, Iowa 50011, USA}
    \affiliation{Department of Materials Science and Engineering, University of Virginia, Charlottesville, Virginia 22904, USA}
  \author{Ross McDonald}
	\affiliation{National High Magnetic Field Laboratory, Los Alamos National Laboratory, Los Alamos, New Mexico 87545, USA}
 \author{Rebecca Flint}
	\affiliation{Ames National Laboratory, U.S. DOE, Iowa State University, Ames, Iowa 50011, USA}
	\affiliation{Department of Physics and Astronomy, Iowa State University, Ames, Iowa 50011, USA}
 \author{ R. J. McQueeney}
	\affiliation{Ames National Laboratory, U.S. DOE, Iowa State University, Ames, Iowa 50011, USA}
	\affiliation{Department of Physics and Astronomy, Iowa State University, Ames, Iowa 50011, USA}
\date{\today}

\begin{abstract}
TbV$_6$Sn$_6$ is a topological metal where ferromagnetic Tb ions with strong uniaxial magnetic anisotropy interact with V kagome layers. Inelastic neutron scattering (INS) measurements show that the Tb ions adopt an Ising doublet ground state. Here, we consider whether a transverse magnetic field can drive TbV$_6$Sn$_6$ toward a quantum critical point, providing a rare example of transverse-field Ising criticality in a metallic compound. High-field magnetization measurements reveal a first-order-like spin-reorientation transition at 25.6 T. Our INS-based magnetic model finds that this is caused by an avoided crossing of an excited-state singlet with the ground-state doublet. Surprisingly, our model predicts that quantum critical and tricritical points are accessible within the range of experimentally determined model parameters and may be reached by varying the direction of an applied magnetic field.
\end{abstract}

\maketitle

    Kagome metals provide a promising platform for discovering topological and flat band physics with recent reports of charge-density wave order \cite{Jiang:2021aa, Teng:2022aa, PhysRevLett.129.216402, PhysRevLett.127.046401}, superconductivity \cite{PhysRevMaterials.5.034801, PhysRevMaterials.3.094407, sciadv.abb6003}, itinerant magnetism \cite{CABLE1993161, sciadv.abe2680, PhysRevB.103.094413, PhysRevB.106.184422, PhysRevB.108.045132}, and Chern insulating phases \cite{Yin:2020aa, Baidya:2019aa, PhysRevLett.126.117602, PhysRevB.80.113102}. Among the many kagome metal systems, $RT_6X_6$ (166) systems with $R=$ rare-earth metal, $T=$ V, Cr, Mn, and $X=$ Sn, Ge are unique for their combination of magnetic $R$ ions that interact with metallic kagome $T$ layers \cite{CHAFIKELIDRISSI1991431, ROMAKA20118862, VENTURINI199299}. 

    Particularly noteworthy is TbV$_6$Sn$_6$: While the vanadium atoms do not generate magnetic moments, the strong spin Berry curvature associated with the kagome Dirac cones leads to a significantly enhanced orbital Zeeman effect \cite{doi:10.1126/sciadv.add2024}. The high-spin Tb moments ($J=6$) order ferromagnetically below $T_{\rm C} = 4.4$~K and have strong Ising anisotropy inferred from low-field magnetization data \cite{PhysRevB.106.115139, PhysRevMaterials.6.105001} and magnetic heat capacity data, which is consistent with an isolated crystal electric field (CEF) doublet ground state \cite{PhysRevMaterials.6.104202}.  Ferromagnetic (FM) order and fluctuations are known to impact transport of conduction electrons in the kagome V layers \cite{PhysRevMaterials.6.104202,PhysRevMaterials.6.105001}. Thus, TbV$_6$Sn$_6$ may be a good candidate for investigating transverse-field Ising quantum criticality and its interaction with kagome electrons.
    
    In the transverse-field Ising model (TFIM) \cite{deGennes}, a transverse magnetic field leads to a quantum critical point (QCP) between Ising FM and quantum paramagnetism. Strong quantum fluctuations near a QCP lead to highly nonclassical behavior and exotic phases of matter \cite{SachdevBook}. There are only a few FM Ising materials, like LiHoF$_4$ \cite{Bitko}, that are described by an effective spin-$1/2$ TFIM up to the critical transverse field, and weakly correlated metallic versions have not been studied.
 
    It is an open question whether TbV$_6$Sn$_6$ will exhibit TFIM quantum criticality where a continuous (second-order) transition is maintained as $T\rightarrow0$. In a high-spin Ising FM, the effect of a transverse field may stabilize other CEF levels that compete with the Ising doublet, leading to a first-order metamagnetic transition \cite{Stasiak}. Even if the doublet remains isolated, the coupling of the Tb moments to the transverse spin fluctuations of conduction electrons may eventually drive the transition first order \cite{Belitz99,Kirkpatrick2012} and generate a quantum critical wing structure, as seen in the heavy fermion URhGe \cite{Yelland2011}.

    In this Letter, we investigate the Ising nature of TbV$_6$Sn$_6$ using inelastic neutron scattering (INS) and high-field magnetization measurements. The INS data allow us to determine the CEF parameters, which characterize a magnetic Hamiltonian with an isolated non-Kramers quasidoublet (Ising) ground state at zero field. To search for evidence of TFIM quantum criticality, we measure the magnetization in a transverse field ($H_{\perp}$) and observe a first-order-like transition at $\mu_0H_{\perp}^c=25.6$~T. Our magnetic model clearly identifies that this metamagnetic (spin-reorientation) transition originates from a CEF level crossing rather than a coupling to conduction electrons.

    However, our model reveals that the true nature of quantum criticality in TbV$_6$Sn$_6$ is subtle and dependent on the strength of the planar magnetic anisotropy, a parameter that is not clearly captured in our experimental data. When the planar anisotropy is small, the QCP is avoided by a first-order transition exhibiting a wing structure with quantum critical end points (QCEPs) occurring in small longitudinal fields.  For larger planar anisotropy, second-order QCPs nucleate along the planar easy axis and eventually expand to all transverse-field directions. In this intermediate regime, TbV$_6$Sn$_6$ is proximate to a realization of a quantum tricritical point (QTCP) that is reachable by varying the magnetic field direction.

\begin{figure*}[] 
    \centering 
	\includegraphics[width=.9\linewidth]{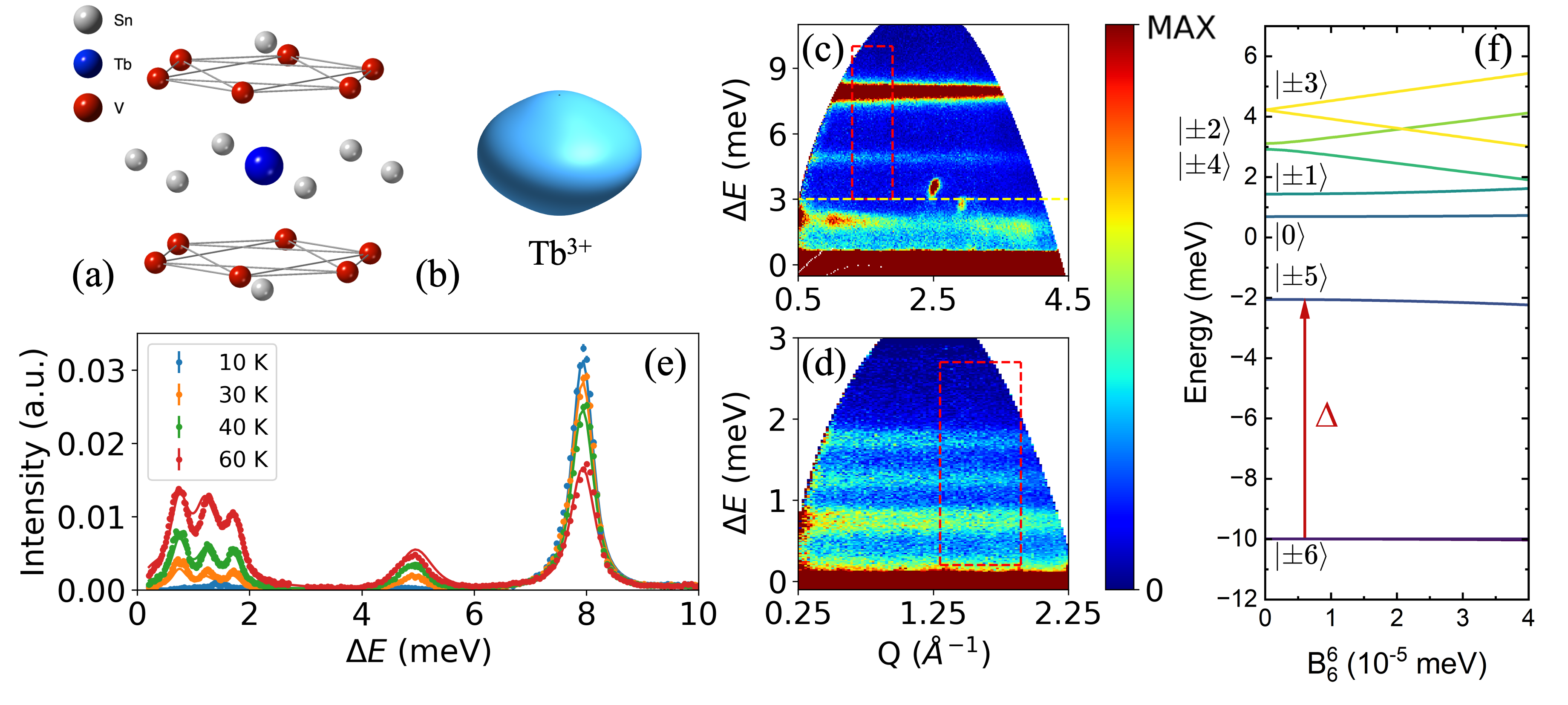}
	\caption{{Ising anisotropy in TbV$_6$Sn$_6$}. The $D_{6h}$ point-group symmetry environment of the Tb$^{3+}$ ion [panel (a)] and its oblate-shaped Tb-4$f$ charge density [panel (b)]. The INS data with incident neutron energy of (c) $E_{i} = 12$~meV and (d) $E_{i} = 3.32$~meV. Data near and below the yellow dashed line in panel (c) are corrupted by spurious instrumental background. (e) The energy dependence of the combined $E_{i} = 3.32$ and 12 meV INS spectra at various temperatures (dots). Intensities at each $E_i$ and $T$ are averaged over the same $Q$ range indicated by the dashed red boxes in panels (c) and (d) and have a fitted background removed. Data for different $E_i$ are placed on the same intensity scale using procedures described in the Supplemental Material \cite{SI}. Fits to the INS spectra from CEF calculations are plotted as solid lines in panel (e). (f) Zero-field CEF energy levels versus $B_6^6$ while keeping other Case I CEF parameters fixed. CEF states are pure $\ket{\pm m}$ eigenstates when $B_6^6=0$.}
    \label{fig:cryst_f} 
\end{figure*} 
{\it Sample preparation}. Single crystals were grown from a Sn flux similar to the previously described process \cite{PhysRevB.106.115139, PhysRevMaterials.6.105001}. An obstacle to growing large TbV$_6$Sn$_6$ crystals, suitable for inelastic neutron scattering, is the relatively poor solubility of V in Sn. We found that the typical reactions described in earlier reports normally yielded mixtures of small TbV$_6$Sn$_6$ crystals alongside chunks of (likely undissolved) V. We empirically found that by using the two-step process described below, larger crystals can be produced.

Elemental pieces of Tb (Ames Lab, 99.99$\%$), V (Alfa-Aesar, 99.99 $\%$) and Sn (Alfa-Aesar, 99.99$\%$) were weighed in a 1:6:60 molar ratio and loaded into the growth side of a 5 ml alumina Canfield crucible set \cite{canfield2016use}. The crucibles were sealed under vacuum in a fused silica ampule and heated in a box furnace to 1180 \textdegree C. After holding at 1180 \textdegree C for $\approx24$ h, the samples were removed from the furnace and the liquid decanted in a metal centrifuge, upon which the liquid phase is cleanly captured in the second crucible on the top side of the crucible set. This first step allows us to cleanly separate the liquid from any undissolved V. A second sequence is then carried out, with the captured decant used in the "growth" side of another crucible set. The second crucible set was sealed under vacuum as before and warmed in a box furnace to 1195 \textdegree C. After holding at 1195 \textdegree C for 6 h, the furnace was cooled to 775 \textdegree C, upon which the ampule was removed from the furnace and the excess liquid decanted. Once cool, the ampules and crucibles were opened to reveal relatively large ($5-10$ mg), single-phase, crystals of TbV$_6$Sn$_6$.

{\it Neutron scattering data}. 
 INS measurements on a polycrystalline sample of TbV$_6$Sn$_6$ were performed on the cold neutron chopper spectrometer at the Spallation Neutron Source at Oak Ridge National Laboratory using incident neutron energies of $E_{i} = 3.32$ and $12$~meV. A polycrystalline sample for INS measurements was prepared by grinding the flux-grown single crystals and sieving with a diameter of 32 $\mathrm{\text\textmu m}$ to obtain particles of uniform size. A sample weighing 1.2 g was sealed in the aluminum can filled with helium gas and mounted to a closed-cycle refrigerator for studies down to $T=1.7$ K. The measured INS intensity is proportional to the spin-spin correlation function $\mathcal{S}(Q, \omega)$, where $Q$ is the momentum transfer, and $\Delta E = \hbar\omega$ is the energy transfer. The data are histogrammed with bin sizes of $\delta E = 0.06$~meV and $\delta Q = 0.02$~\AA$^{-1}$ using $\text{DAVE}$ software \cite{DAVE}.

    Figures~\ref{fig:cryst_f}(c) and (d) show the INS data at $T=30$ K, well above $T_\text{C} = 4.4$~K, with  $E_{i} = 12$~ and 3.32 meV respectively. These data show multiple non-dispersive excitations between  CEF levels of the Tb$^{3+}$ ion. The $Q$-averaged intensities are plotted versus energy in Fig.~\ref{fig:cryst_f}(e) for $T = 10$, $30$, $40$, and $60$~K. The main transition out of the ground state appears at $\Delta E = 7.9$~meV at $T = 10$~K and its intensity drops as the ground state thermally depopulates when raising the temperature. Multiple transitions occur nearly concurrently above 30~K suggesting that excited CEF levels are well separated from the ground-state doublet, but lie close to each other in energy.

    Detailed information about the CEF can be obtained by fitting the INS spectra. The Hamiltonian for a Tb$^{3+}$ ion in a hexagonal CEF with $D_{6h}$ point-group symmetry [see Fig.~\ref{fig:cryst_f}(a)] is 
\begin{eqnarray}
\mathcal{H}_{\text{CEF}} = B_2^0\mathcal{O}_2^0 +B_4^0\mathcal{O}_4^0 +B_6^0\mathcal{O}_6^0 +B_6^6\mathcal{O}_6^6.
\label{eq:CEF Hamiltonian}
\end{eqnarray}
    We assume the Tb ion is in its Hund's rule ground state with total angular momentum $J=6$ ($S=3$, $L=3$, and $g_J = 3/2$). $\mathcal{O}_l^m$ are the corresponding Stevens operator equivalents for $J=6$ and $B_l^m$ are the CEF parameters.

    For the measurement of CEF excitations in powder samples with unpolarized neutrons, the neutron intensity is given by
\begin{equation}
\mathcal{S}(Q,\omega) \propto \sum_{j,k}\sum_\alpha p_j|\bra{\Gamma_k}J_\alpha\ket{\Gamma_j}|^2 \delta(E_k-E_j-\hbar\omega).
\label{eqn:neutron}
\end{equation}
    Here, $j$ labels an initial CEF state $\ket{\Gamma_j}$ with energy $E_j$ and thermal population $p_j$ and $k$ labels a final state $\ket{\Gamma_k}$ with energy $E_k$. INS transitions follow the dipole selection rules for $J_{\alpha=x,y,z}$ matrix elements ($\Delta m = 0,\pm1$, where $m$ is the magnetic quantum number).

    Using both the $\text{PyCrystalField}$ and $\text{LMFIT}$ packages \cite{Scheie:in5044,LMFIT}, we simultaneously fit INS spectra at all temperatures in Fig.~\ref{fig:cryst_f}(e) using Eqns.~(\ref{eq:CEF Hamiltonian}) and (\ref{eqn:neutron}). We use the Voigt peak profile convoluting the resolution and intrinsic widths since the CEF peaks are broader than the instrumental resolution. We achieved two reasonable CEF parameter sets labeled Cases I and II, as shown in Table \ref{tbl:CEF_parameters} and described in detail in the Supplemental Material \cite{SI}. The main difference is the size of the $B_6^6$ parameter, which is nearly an order of magnitude larger in Case II. This parameter establishes the planar magnetic anisotropy and its magnitude is pivotal in the development of quantum critical behavior, as described below.

\begin{table}[]
\caption{Crystal-field parameters for $\text{TbV}_6\text{Sn}_6$ (in meV).}
\begin{tabular}{c|ccccc}
\hline\hline
Case & $B_2^0$ & $B_4^0~(\times10^4)$ & $B_6^0~(\times10^6)$ & $B_6^6~(\times10^6)$ & $\chi_{\nu}^2$ \\
\hline 
I & $-0.1036(4)$ & $-5.91(2)$ & $2.31(2)$ & $3.3(6)$ & $8.6$ \\
II & $-0.1151(3)$ & $-5.57(1)$ & $2.31(2)$ & $23.0(5)$ & $11.6$ \\
\hline\hline
\end{tabular}
\label{tbl:CEF_parameters}
\end{table}

    Figure \ref{fig:cryst_f}(f) shows the calculated CEF levels in Case I for variable $B_6^6$. For small $B_6^6$, the CEF eigenstates can be labeled by $\ket{\Gamma_j} \approx \ket{m}$ and the $\mathcal{O}_6^6$ operator causes small mixing between $\ket{m}$ and $\ket{m\pm6}$ states (the eigenvectors are listed in Table~S1 of the Supplemental Material \cite{SI}). As expected, we find that the ground state is an Ising doublet $\ket{\pm 6}$ with a strongly anisotropic orbital wave function shown in Fig.~\ref{fig:cryst_f}(b). In both cases, the ground state doublet is well separated from the first excited state $\approx \ket{\pm 5}$ by $\Delta=$ 7.9 meV with the ratio $k_{\rm B}T_\text{C}/\Delta\approx0.05$, consistent with an  Ising character of the FM ordered state. Thermodynamic calculations from these CEF models agree with magnetization (see below), susceptibility, and heat capacity data (see Fig. S1 in the Supplemental Material \cite{SI}).

\begin{figure*}[]
\includegraphics[width=.9\linewidth]{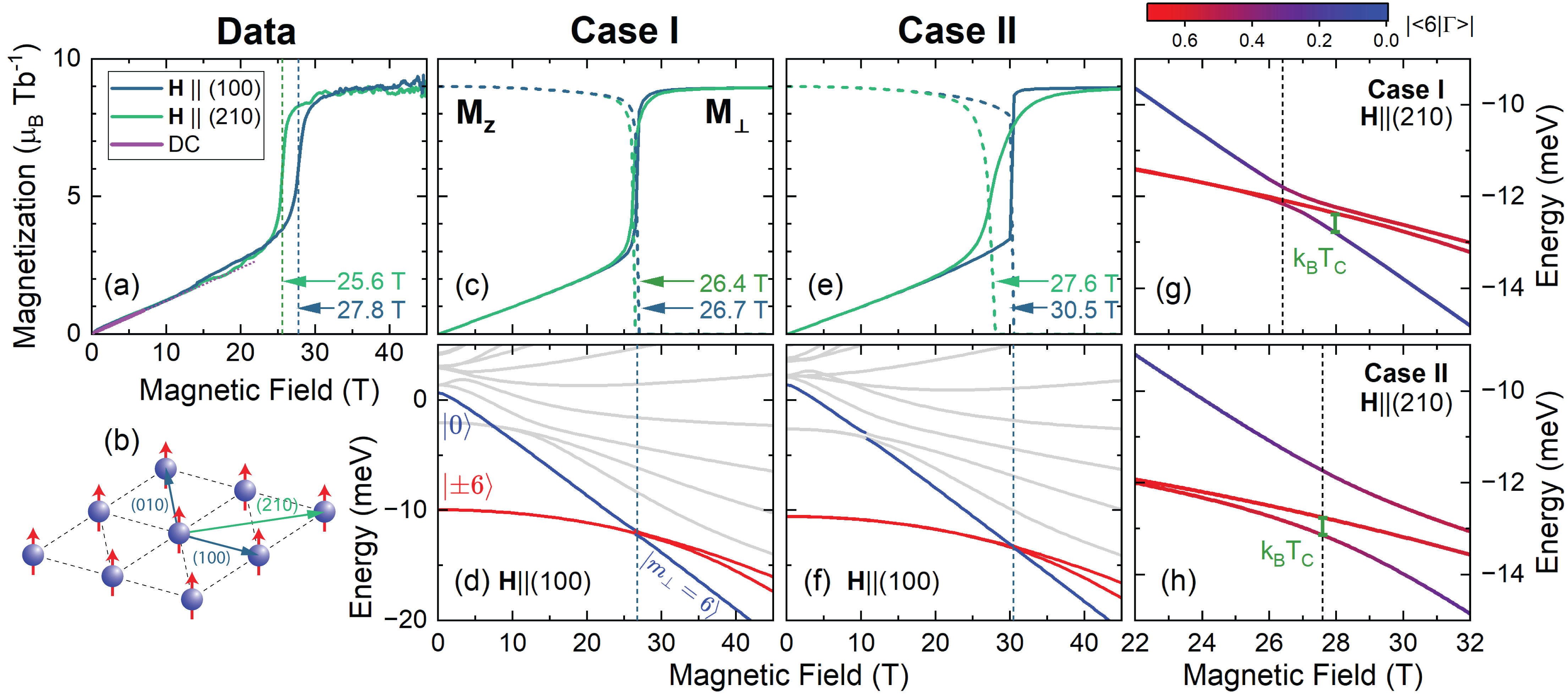}
	\caption{{Transverse field quantum phase transition}. (a) Magnetization data measured in a pulsed transverse field along (100) and (210) directions at $T=625$ mK. Low-field dd data measured at 1.8 K are shown for comparison. (b) Triangular Tb layer showing field directions. (c) Case I mean-field model calculations of the magnetization components parallel to the (100) and (210) applied field directions ($M_{\perp}$, solid lines) and along (001) ($M_z$, dashed lines). (d) Evolution of crystal-field levels with transverse field applied along (100). The Ising state doublet (red lines) and the state that evolves from the zero-field $\ket{m=0}$ state to the planar $\ket{m_{\perp}=6}$ state (blue line) are highlighted. (e), (f) Mean-field calculations for Case II parameters. Panels (g) and (h) show the evolution of the Case I and Case II CEF levels with transverse field along (210). Lines are colored according to projection of their wavefunction $\ket{\Gamma}$ onto the $\ket{6}$ state. Dashed lines show the critical field and the green bar is $\sim k_{\rm B} T_{\rm C}$.}
    \label{fig:mag}
\end{figure*}

{\it High-field magnetization data}. 
    To search for evidence of TFIM quantum criticality, we performed magnetization measurements on TbV$_6$Sn$_6$ in pulsed magnetic fields up to 60~T at National High Magnetic Field Laboratory at Los Alamos National Laboratory using a compensated coil extraction magnetometer. The voltage induced in the coil by the field pulse is proportional to the sample susceptibility ($dM/dH$). Magnetization [$M(H)$] data are obtained by subtracting the empty coil signal from the data with the sample inside the coil and integrating with respect to the field (time). Two different crystal mountings were investigated with the field parallel to inequivalent (100) or (210) high-symmetry directions in the hexagonal plane, as shown in Fig.~\ref{fig:mag}(b).  Figure \ref{fig:mag}(a) compares the magnetization with field applied in either direction (raw $dM/dH$ data are shown in Fig.~S3 os the Supplemental Material \cite{SI}). The magnetization initially tilts weakly toward the field due to the strong Ising anisotropy and then jumps parallel to the field in a first-order-like spin-reorientation transition. We observe critical fields of $25.6$ and $27.8$ T for (210) and (100) field directions, respectively,  suggesting a planar easy axis along the (210) direction ($B_6^6>0$). The first-order-like character of the spin-reorientation transition provides clues to the nature of the quantum criticality in TbV$_6$Sn$_6$ that are revealed by theoretical analysis described below.

    Independent estimates of the $B_l^0$ CEF parameters derived from magnetization data are described in the Supplemental Material \cite{SI} and provide values similar to INS data. The observed critical field anisotropy of 2.2 T between the two field directions  allow for an independent determination  $B_6^6 \approx 30\times 10^{-6}$~meV in support of Case II. However, high-field proximity detector oscillator (PDO) measurements described in the SI \cite{SI} find evidence that the sample orientation is tilted by strong torques. PDO measurements were performed while varying the in-plane field direction and found a minimum critical field of 25.6 T for ${\bf H}\parallel(210)$, but also found a prominent twofold rotational component, rather than the expected sixfold pattern, which suggests sample misalignment. Tilting introduces a small longitudinal component of the applied field [along the (001) axis] that increases the apparent critical field by approximately $1$~T per degree of tilt (see SI \cite{SI}). Thus, the measured critical field anisotropy is caused by misalignment and we cannot distinguish Case I from Case II.
    
{\it Quantum criticality}.

\begin{figure*}[htbp]
    \centering 
	\includegraphics[width=.95\linewidth]{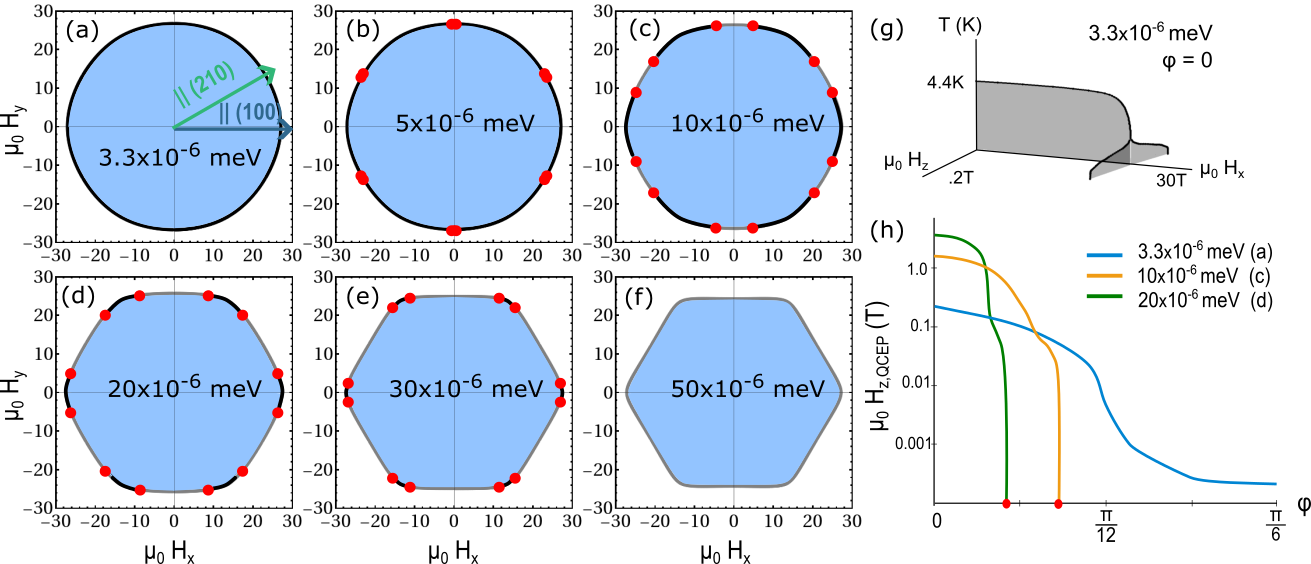}
	\caption{{Quantum criticality in TbV$_6$Sn$_6$}.  
    (a)$-$(f) Mean-field phase diagrams in the $H_x$-$H_y$ plane [with $(100)$ and $(210)$ directions shown in panel (a)] for Case I CEF parameters with variable $B_6^6$ values; the $B_l^0$ and $\mathcal{J}_0$ are held fixed. The blue regions are ferromagnetic and the white regions are quantum paramagnetic. There are two critical $B_6^6$ values: a lower $B_6^6 = 4.6\times 10^{-6}$meV, below which all transitions are first order (black), and an upper $B_6^6= 33\times 10^{-6}$meV, above which all transitions are second order (gray).  In-between, there are QTCPs at certain angles (red dots) that separate the first- and second-order lines.  Case II behaves very similarly to $B_6^6 = 20\times 10^{-6}$meV.  (g) Quantum critical wing structure in $H_x$-$H_z$ plane for Case I parameters. (h) Evolution of the angular dependence of the QCEPs with $B_6^6$, in the $H_x$-$H_y$ plane. The red dots indicate where the QCEPs evolve into QTCPs for larger $B_6^6$.
    }
    \label{fig:theory}
\end{figure*} 

    To understand the quantum critical behavior of TbV$_6$Sn$_6$, mean-field (MF) calculations of the magnetic Hamiltonian were performed. We extend the CEF Hamiltonian in Eq.~(\ref{eq:CEF Hamiltonian}) to include an applied magnetic field and an effective isotropic FM Heisenberg coupling between Tb ions with total angular momentum $J$,
\begin{align}
\label{eq:ham}
\mathcal{H} =  \mathcal{H}_{\text{CEF}}
-\frac{1}{2} \sum_{ij} \mathcal{J}_{ij} \vec{J}_i \cdot \vec{J}_j
- g_J \mu_B \mu_0\vec{H} \cdot \sum_i \vec{J}_i.
\end{align}
    Hyperfine corrections and long-range dipolar interactions are estimated to be small, as discussed in the Supplemental Material \cite{SI}.  The MF transition temperature is set by $\mathcal{J}_0 = \sum_j \mathcal{J}_{ij}$ and we choose $\mathcal{J}_0 = 122$~mK to reproduce the experimental $T_{\rm C} = 4.4$~K. Fluctuations will decrease the critical temperature, making this value a lower bound of the interaction strength. 
    
    We first employ  the Case I CEF parameters where planar anisotropy is small and find that the zero-temperature quantum phase transition is first order for all in-plane field directions indicating that the TFIM QCP is avoided. The calculated magnetization, $\vec{M} = g_J \langle \vec{J}\rangle$, as a function of transverse field is shown in Fig.~\ref{fig:mag}(c), where the critical fields are weakly anisotropic: $26.4$~T along the (210) easy axis and $26.7$~T along (100). Figure~\ref{fig:mag}(d) shows that the first-order transition is a metamagnetic spin reorientation triggered by a CEF level crossing in the transverse field. Near the critical field, the original ground-state doublet (red lines) mixes with the original $|0\rangle$ state  (blue line) that evolves into the $\approx |m_\perp = 6\rangle$ state appropriate for a field-aligned planar moment.

    According to our MF calculations, the spin-reorientation transition becomes second order at finite temperature, with a classical tricritical point at $T^{*} =1.65$ K (1 K) along the easy (hard) axis, respectively. The tricritical point leads to a quantum critical wing structure \cite{Griffiths} in small $c$-axis fields. This wing structure is very compact, with quantum critical end points at $\mu_0(H_x, H_z) \approx (27,\pm 0.2)$ T along the hard axis, as shown in Figs. \ref{fig:theory}(g) and \ref{fig:theory}(h). Similar to a QCP, quantum critical endpoints possess quantum fluctuations that may mediate exotic high-field superconductivity observed in URhGe \cite{Levy, Huxley}and UTe$_2$ \cite{aoki2022unconventional,PhysRevX.15.021019,miyake2019metamagnetic,sundar2019coexistence,duan2021resonance}.

    Calculations using Case II CEF parameters [Fig.~\ref{fig:mag}(e)] provide a more interesting scenario where a TFIM quantum critical (second-order) transition is observed for the field along the easy axis, while a first-order transition is observed for the hard axis. This indicates that a quantum tricritical point separating first- and second-order transitions would be accessible by varying the field direction.  Close to the hard axis, Fig.~\ref{fig:theory}(h) shows that a quantum critical endpoint appears for fields tilted out of the plane.
    Figures~\ref{fig:mag}(f) and \ref{fig:mag}(h) show that quantum critical behavior in Case II is influenced by the same level crossing. 

    By comparing the two cases, we find that the nature of quantum criticality in TbV$_6$Sn$_6$ is highly sensitive to the $B_6^6$ CEF parameter. To illustrate this, we study how the quantum criticality is tuned by $B_6^6$ with other Case I parameters held fixed.  The mean-field ground-state phase diagrams in atransverse field are shown in Fig.~\ref{fig:theory}(a)$-$(f) for evolving $B_6^6$, where we find two critical values of $B_6^6$. Below the lower critical $B_6^6 = 4.6\times 10^{-6}$~meV, the transverse-field quantum transitions are all first-order, but above it, a line of second-order transitions nucleates around the (210) easy axis, separated from the first-order transitions by QTCPs at finite field angles. As $B_6^6$ increases, the line of second-order QCPs grows until the QTCPs disappear at the upper critical $B_6^6 = 33\times 10^{-6}$~meV, above which the transition is second order for all in-plane field directions. The phase diagram evolves from circular to hexagonal as the planar anisotropy $B_6^6$ increases.

    The conditions for quantum critical behavior are understood by examining Fig.~\ref{fig:mag}(g) and \ref{fig:mag}(h) that show the evolution of CEF levels near the critical field for Cases I and II that, respectively, host first-order and second-order transitions. In the spin-1/2 Ising model, a QCP occurs when the splitting of the doublet is large enough to overcome the longitudinal interaction strength ($\sim k_{\rm B}T_{\rm C}$) \cite{SuzukiBook}. For small $B_6^6$, Fig.~\ref{fig:mag}(g) shows that a weakly avoided level crossing with the excited CEF singlet occurs before the doublet splitting is large enough to overcome the interaction strength, resulting in a first-order metamagnetic transition. In Fig.~\ref{fig:mag}(h), the larger value of $B_6^6$ increases level repulsion between the Ising doublet and the excited singlet. The splitting of the Ising doublet is sufficient to overcome the interaction without strong mixing between the doublet and excited singlet, leading to a continuous transition.  The field direction dependence of the doublet splitting then allows for the development of quantum critical end points and tricritical points (see Supplemental Material for details \cite{SI}).

{\it Discussion}.
    We have shown that the kagome metal TbV$_6$Sn$_6$ is a good realization of an Ising ferromagnet. Remarkably, whether or not it exhibits TFIM quantum criticality, quantum tricriticality or a wing structure with quantum critical end points depends sensitively on the crystal fields and the angle of the transverse field in the $ab$ plane. This discovery reveals a rare opportunity to address the nature and tunability of quantum critical Ising behavior in a non-heavy-fermion metal.

    However, our measurements show that the true nature of quantum criticality in TbV$_6$Sn$_6$ is somewhat elusive. Our best set of CEF parameters (Case I) places TbV$_6$Sn$_6$ below the lower critical value of $B_6^6=4.6\times10^{-6}$ meV.  The resulting first-order metamagnetic transition in a transverse field is predicted to host a wing structure with quantum critical end points in a tilted field. In this respect, the critical behavior is reminiscent of the Blume-Capel model describing an anisotropic spin-1 FM where crossing of the non-magnetic singlet with the doublet controls the quantum criticality \cite{PhysRev.141.517, capel1966possibility}.
    
    Much larger values of $B_6^6$ (Case II) are also consistent with the experimental data and would put TbV$_6$Sn$_6$ in a regime where the quantum critical nature varies with the in-plane field angle, including QTCPs at particular angles. Density-functional theory estimates ($B_6^6 = 21 \times 10^{-6}$ meV, see the Supplemental Material \cite{SI}) are consistent with $B_6^6$ values in this interesting quantum critical regime.  We note that larger values of $B_6^6$ will not affect the Ising character of TbV$_6$Sn$_6$, as shown in Fig.~\ref{fig:cryst_f}(f). 

    Thus, our analysis places TbV$_6$Sn$_6$ within the reach of field-angle-tuned quantum critical and tricritical behavior. Proximity to QTCPs has been suggested to explain unusual behavior in Sr$_3$Ru$_2$O$_7$ \cite{Grigera,Green,Mercaldo}, URhGe \cite{Levy,Huxley}, YbRh$_2$Si$_2$, and other materials \cite{Misawa2008,Giovannetti,Friedemann2018,Kaluarachchi2018, Ullah2024}. The capability to tune $RT_6X_6$ materials themselves (by chemical substitution, for example) to increase the planar anisotropy could provide a more definitive example of quantum tricriticality.

    The fields required to tune TbV$_6$Sn$_6$ into this quantum critical regime are large, but experimentally accessible, making it a potential testing ground to compare the signatures of quantum criticality or tricriticality by tuning the angle of the in-plane magnetic field.  However, sample tilting in the presence of large torques would need to be very carefully avoided.  One interesting possibility is to search for the finite temperature signatures of quantum criticality.  For example, the longitudinal magnetic susceptibility should diverge with different power laws above the QTCPs, $\chi_{zz} \sim T^{-2\psi}$, compared to the QCPs, $\chi_{zz} \sim T^{-\psi}$ \cite{Mercaldo,Kato2015}, where $\psi$ is likely $2$ for short-range three-dimensional interactions. Alternately, high-field transport measurements could reveal that the critical points are masked by the formation of low-temperature phases, as with the superconductivity in URhGe \cite{Levy, Huxley} or UTe$_2$ \cite{aoki2022unconventional,PhysRevX.15.021019,miyake2019metamagnetic,sundar2019coexistence,duan2021resonance}, and the nematic spin density waves in ultrapure Sr$_3$Ru$_2$O$_7$ \cite{Green, Lester}.

\begin{acknowledgments}
{\it Acknowledgments}.
The work of T.H., B.L., and R.J.M. (inelastic neutron scattering), R.D.M. and R.F. (mean-field analysis), and Y.L. and L.K. (density-functional theory) at the Ames National Laboratory was supported by the U.S. Department of Energy, Office of Science, Basic Energy Sciences, Materials Science and Engineering. The work of J.B. and R.M. (high-field measurements) T.J.S. and P.C.C. (crystal growth and characterizations) was supported by the Center for the Advancement of Topological Semimetals (CATS), an Energy Frontier Research Center funded by the  U.S. Department of Energy (DOE) Office of Science (SC), Office of Basic Energy Sciences (BES), through the Ames National Laboratory. Ames National Laboratory is operated for the U.S. DOE by Iowa State University under Contract No. DE-AC02-07CH11358. This research used resources at the Spallation Neutron Source, a DOE Office of Science User Facility operated by the Oak Ridge National Laboratory. A portion of this work was performed at the National High Magnetic Field Laboratory, which is supported by National Science Foundation Cooperative Agreement No. DMR-2128556, the U.S. DOE, and the state of Florida.

{\it Data avalibility}. 
The data that support the findings of this article are openly available\cite{dataAvaliable}.
\end{acknowledgments}

\bibliography{TbV6Sn6Bib}

\clearpage
\raggedright
\end{document}


\setcounter{equation}{0}
	\setcounter{figure}{0}
	\setcounter{table}{0}
	\setcounter{page}{1}
	\setcounter{section}{0}
	\makeatletter
    
	\renewcommand{\theequation}{S\arabic{equation}}
	\renewcommand{\thefigure}{S\arabic{figure}}
	\renewcommand{\thetable}{S\arabic{table}}
 	\renewcommand{\thesection}{S\arabic{section}}
	\renewcommand{\bibnumfmt}[1]{[S#1]}
	\renewcommand{\citenumfont}[1]{S#1}

\title{Supplementary Information - Nature of quantum criticality in the Ising ferromagnet TbV$_{\bm{6}}$Sn$_{\bm{6}}$}
\author{Tianxiong~Han}
	\affiliation{Ames National Laboratory, U.S. DOE, Iowa State University, Ames, Iowa 50011, USA}
	\affiliation{Department of Physics and Astronomy, Iowa State University, Ames, Iowa 50011, USA}
 \author{R. D. McKenzie}
	\affiliation{Ames National Laboratory, U.S. DOE, Iowa State University, Ames, Iowa 50011, USA}
	\affiliation{Department of Physics and Astronomy, Iowa State University, Ames, Iowa 50011, USA}
 \author{Joanna Blawat}
	\affiliation{National High Magnetic Field Laboratory, Los Alamos National Laboratory, Los Alamos, NM 87545}
 \author{Tyler J. Slade}
	\affiliation{Ames National Laboratory, U.S. DOE, Iowa State University, Ames, Iowa 50011, USA}
  \author{Bing Li}
  	\affiliation{Ames National Laboratory, U.S. DOE, Iowa State University, Ames, Iowa 50011, USA}
    \affiliation{Oak Ridge National Laboratory, Oak Ridge, TN, 37831, USA}
  \author{Y. Lee}
	\affiliation{Ames National Laboratory, U.S. DOE, Iowa State University, Ames, Iowa 50011, USA}
 \author{D. M. Pajerowski}
\affiliation{Oak Ridge National Laboratory, Oak Ridge, TN, 37831, USA}
 \author{John Singleton}
	\affiliation{National High Magnetic Field Laboratory, Los Alamos National Laboratory, Los Alamos, NM 87545}
 \author{Paul C. Canfield}
	\affiliation{Ames National Laboratory, U.S. DOE, Iowa State University, Ames, Iowa 50011, USA}
	\affiliation{Department of Physics and Astronomy, Iowa State University, Ames, Iowa 50011, USA}
  \author{Liqin Ke}
	\affiliation{Ames National Laboratory, U.S. DOE, Iowa State University, Ames, Iowa 50011, USA}
    \affiliation{Department of Materials Science and Engineering, University of Virginia, Charlottesville, VA 22904}
  \author{Ross McDonald}
	\affiliation{National High Magnetic Field Laboratory, Los Alamos National Laboratory, Los Alamos, NM 87545}
 \author{Rebecca Flint}
	\affiliation{Ames National Laboratory, U.S. DOE, Iowa State University, Ames, Iowa 50011, USA}
	\affiliation{Department of Physics and Astronomy, Iowa State University, Ames, Iowa 50011, USA}
 \author{ R. J. McQueeney}
	\affiliation{Ames National Laboratory, U.S. DOE, Iowa State University, Ames, Iowa 50011, USA}
	\affiliation{Department of Physics and Astronomy, Iowa State University, Ames, Iowa 50011, USA}
\date{\today}

\maketitle


\section{Extended neutron data analysis}
\label{ap:CaseII}
\subsection{Eigenvalues and eigenvectors for CEF Hamiltonian}

After diagonalizing the Hamiltonian using both sets of CEF parameters, we obtain the zero-field the eigenvalues and eigenvectors, as shown in Table~\ref{tbl:wave function}.

\begin{table*}
\caption{\textbf{Crystal-field eigenvalues and wavefunctions for TbV$_\textbf{6}$Sn$_\textbf{6}$}.}
\begin{ruledtabular}
\begin{tabular}{c|ccccccccccccc}
E (meV) &$|-6\rangle$ & $|-5\rangle$ & $|-4\rangle$ & $|-3\rangle$ & $|-2\rangle$ & $|-1\rangle$ & $|0\rangle$ & $|1\rangle$ & $|2\rangle$ & $|3\rangle$ & $|4\rangle$ & $|5\rangle$ & $|6\rangle$ \tabularnewline
\hline 
\multicolumn{14}{c}{Case I}\\
\hline 
0.000 & 0.7071 & 0.0 & 0.0 & 0.0 & 0.0 & 0.0 & 0.0048 & 0.0 & 0.0 & 0.0 & 0.0 & 0.0 & 0.7071 \tabularnewline
0.000 & -0.7071 & 0.0 & 0.0 & 0.0 & 0.0 & 0.0 & -0.0 & 0.0 & 0.0 & 0.0 & 0.0 & 0.0 & 0.7071 \tabularnewline
7.937 & 0.0 & -0.9998 & 0.0 & 0.0 & 0.0 & 0.0 & 0.0 & -0.0194 & 0.0 & 0.0 & 0.0 & 0.0 & 0.0 \tabularnewline
7.937 & 0.0 & 0.0 & 0.0 & 0.0 & 0.0 & 0.0194 & 0.0 & 0.0 & 0.0 & 0.0 & 0.0 & 0.9998 & 0.0 \tabularnewline
10.683 & -0.0034 & 0.0 & 0.0 & 0.0 & 0.0 & 0.0 & 1.0 & 0.0 & 0.0 & 0.0 & 0.0 & 0.0 & -0.0034 \tabularnewline
11.431 & 0.0 & -0.0194 & 0.0 & 0.0 & 0.0 & 0.0 & 0.0 & 0.9998 & 0.0 & 0.0 & 0.0 & 0.0 & 0.0 \tabularnewline
11.431 & 0.0 & 0.0 & 0.0 & 0.0 & 0.0 & 0.9998 & 0.0 & 0.0 & 0.0 & 0.0 & 0.0 & -0.0194 & 0.0 \tabularnewline
12.877 & 0.0 & 0.0 & 0.0 & 0.0 & -0.3674 & 0.0 & 0.0 & 0.0 & 0.0 & 0.0 & -0.93 & 0.0 & 0.0 \tabularnewline
12.877 & 0.0 & 0.0 & 0.93 & 0.0 & 0.0 & 0.0 & 0.0 & 0.0 & 0.3674 & 0.0 & 0.0 & 0.0 & 0.0 \tabularnewline
13.144 & 0.0 & 0.0 & 0.0 & 0.0 & 0.93 & 0.0 & 0.0 & 0.0 & 0.0 & 0.0 & -0.3674 & 0.0 & 0.0 \tabularnewline
13.144 & 0.0 & 0.0 & -0.3674 & 0.0 & 0.0 & 0.0 & 0.0 & 0.0 & 0.93 & 0.0 & 0.0 & 0.0 & 0.0 \tabularnewline
14.120 & 0.0 & 0.0 & 0.0 & -0.7071 & 0.0 & 0.0 & 0.0 & 0.0 & 0.0 & -0.7071 & 0.0 & 0.0 & 0.0 \tabularnewline
14.320 & 0.0 & 0.0 & 0.0 & 0.7071 & 0.0 & 0.0 & 0.0 & 0.0 & 0.0 & -0.7071 & 0.0 & 0.0 & 0.0 \tabularnewline
 \hline 
 \multicolumn{14}{c}{Case II}\\
 \hline
0.000 & 0.7068 & 0.0 & 0.0 & 0.0 & 0.0 & 0.0 & -0.0293 & 0.0 & 0.0 & 0.0 & 0.0 & 0.0 & 0.7068 \tabularnewline
0.010 & -0.7071 & 0.0 & 0.0 & 0.0 & 0.0 & 0.0 & -0.0 & 0.0 & 0.0 & 0.0 & 0.0 & 0.0 & 0.7071 \tabularnewline
7.932 & 0.0 & -0.995 & 0.0 & 0.0 & 0.0 & 0.0 & 0.0 & 0.0996 & 0.0 & 0.0 & 0.0 & 0.0 & 0.0 \tabularnewline
7.932 & 0.0 & 0.0 & 0.0 & 0.0 & 0.0 & -0.0996 & 0.0 & 0.0 & 0.0 & 0.0 & 0.0 & 0.995 & 0.0 \tabularnewline
11.870 & -0.0208 & 0.0 & 0.0 & 0.0 & 0.0 & 0.0 & -0.9996 & 0.0 & 0.0 & 0.0 & 0.0 & 0.0 & -0.0208 \tabularnewline
12.578 & 0.0 & -0.0996 & 0.0 & 0.0 & 0.0 & 0.0 & 0.0 & -0.995 & 0.0 & 0.0 & 0.0 & 0.0 & 0.0 \tabularnewline
12.578 & 0.0 & 0.0 & 0.0 & 0.0 & 0.0 & -0.995 & 0.0 & 0.0 & 0.0 & 0.0 & 0.0 & -0.0996 & 0.0 \tabularnewline
12.853 & 0.0 & 0.0 & 0.0 & 0.0 & -0.4758 & 0.0 & 0.0 & 0.0 & 0.0 & 0.0 & 0.8796 & 0.0 & 0.0 \tabularnewline
12.853 & 0.0 & 0.0 & 0.8796 & 0.0 & 0.0 & 0.0 & 0.0 & 0.0 & -0.4758 & 0.0 & 0.0 & 0.0 & 0.0 \tabularnewline
14.133 & 0.0 & 0.0 & 0.0 & -0.7071 & 0.0 & 0.0 & 0.0 & 0.0 & 0.0 & 0.7071 & 0.0 & 0.0 & 0.0 \tabularnewline
14.337 & 0.0 & 0.0 & -0.4758 & 0.0 & 0.0 & 0.0 & 0.0 & 0.0 & -0.8796 & 0.0 & 0.0 & 0.0 & 0.0 \tabularnewline
14.337 & 0.0 & 0.0 & 0.0 & 0.0 & 0.8796 & 0.0 & 0.0 & 0.0 & 0.0 & 0.0 & 0.4758 & 0.0 & 0.0 \tabularnewline
15.494 & 0.0 & 0.0 & 0.0 & 0.7071 & 0.0 & 0.0 & 0.0 & 0.0 & 0.0 & 0.7071 & 0.0 & 0.0 & 0.0 \tabularnewline
\end{tabular}\end{ruledtabular}
\label{tbl:wave function}
\end{table*}

\subsection{ Case II CEF model}
The INS data can be adequately represented by larger $B_6^6$ values. The reduced $\chi_\nu^2$ map as a function of $B_6^6$ and the intensity scaling ratio between $E_i =$ 12 and 3.3 meV data is shown in {Fig.~\ref{fig:chisq}(a). Whereas the global minimum that provides best fit is  Case I, the Case II parameters with much larger $B_6^6$ (and larger scaling ratio) gives a local minimum. Using the incoherent elastic line, we independently verified that the scaling ratio is in the range of $1$ to $1.3$ corresponding to the Case I global minimum.

The local minimum of Case II has CEF parameters with $B_2^0 = -0.1151(3)$, $B_4^0 = -5.57(1)\times10^{-4}$, $B_6^0 = 2.31(2)\times10^{-6}$, $B_6^6=2.30(5)\times10^{-5}$ meV, and $\chi_\nu^2 = 11.64$ for comparison. The calculated neutron spectrum from Case II parameters compared to the measurement is shown in {Fig.~\ref{fig:chisq}(b). We note that the larger $B_6^6$ term more thoroughly mixes the $\ket{\pm m}$ CEF states which results in additional dipole-allowed transitions in the energy spectrum as shown in the inset of \ref{fig:chisq}(b).
\begin{figure}
    \centering
    \includegraphics[width=\linewidth]{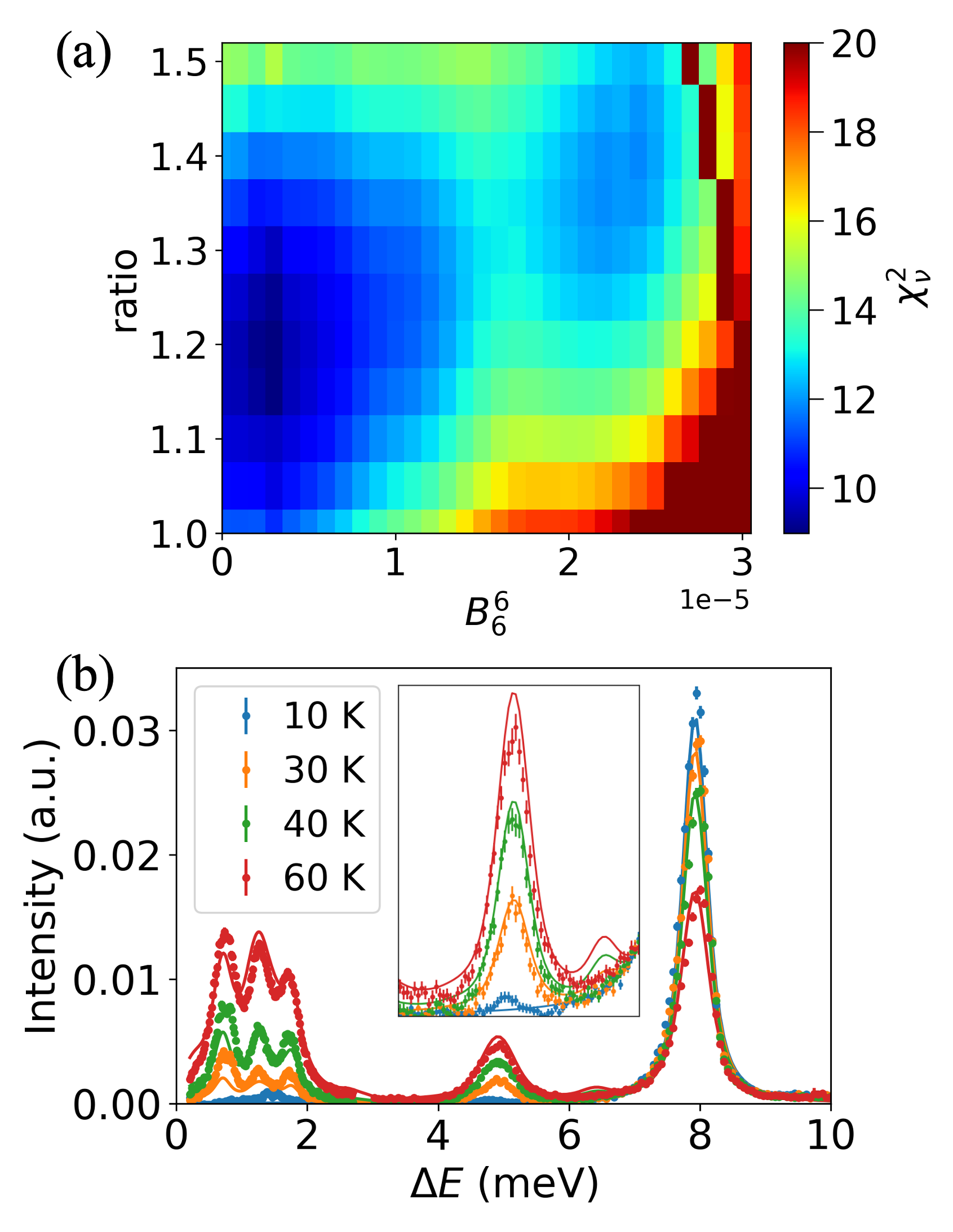}
    \caption{ (a) Reduced $\chi_\nu^2$ map as a function of scaling ratio and $B_6^6$ values that are fixed while fitting other CEF parameters. The global minimum (Case I) is found with ratio $R = 1.2$ and $B_6^6 = 3.3\times10^{-6}$ meV, and a local minimum (Case II) is found with $R = 1.4$ and $B_6^6 = 2.3\times10^{-5}$ meV. (b) The Case II parameters show the extra peaks in the spectrum. The insert shows the zoom-in between $4$~meV$<\Delta E <7$~meV where Case II parameters predict an extra excitation that is not observed from INS measurements.}
    \label{fig:chisq}
\end{figure}

Using these parameters, one finds a continuous transition at $27.6$ T along the easy (210) axis, and a first order transition at $30.7$ T along the hard (100) axis. A quantum tricritical point (QTCP) occurs as the transverse field is rotated in the basal plane. If we take $\phi=0^{\circ}$ to correspond to the hard (100) axis the QTCP occurs at $\phi=5.5^{\circ}$.

\section{Extended magnetization data analysis}

Low field ($\mu_0 H =$ 0--7 T) magnetization measurements were performed in a Quantum Design Magnetic Property Measurement System (MPMS-classic) SQUID magnetometer operating in the DC measurement mode. For measurements with ${\bf H} \parallel$ (100), the sample was mounted between two straws. For measurements with ${\bf H} \parallel$ \textit{c}, the sample was glued to a Kel-F disc. In the latter case, a blank reference sequence over the same temperature and fields was measured on the bare disc and used for a background subtraction. The temperature dependent measurements were field cooled starting at 300 K, and the magnetization isotherms measured on decreasing field from 7 T.  Transverse field data are shown in Fig.~3(a) in the main text.

\subsection{Model-based thermodynamics calculations}
Fig.~\ref{fig:thermo}(a) shows measurements of the anisotropic magnetic susceptibility compared to MF calculations. Calculations of the magnetic heat capacity and entropy are shown in Fig.~\ref{fig:thermo}(b)].

\begin{figure}
    \centering
    \includegraphics[width=\linewidth]{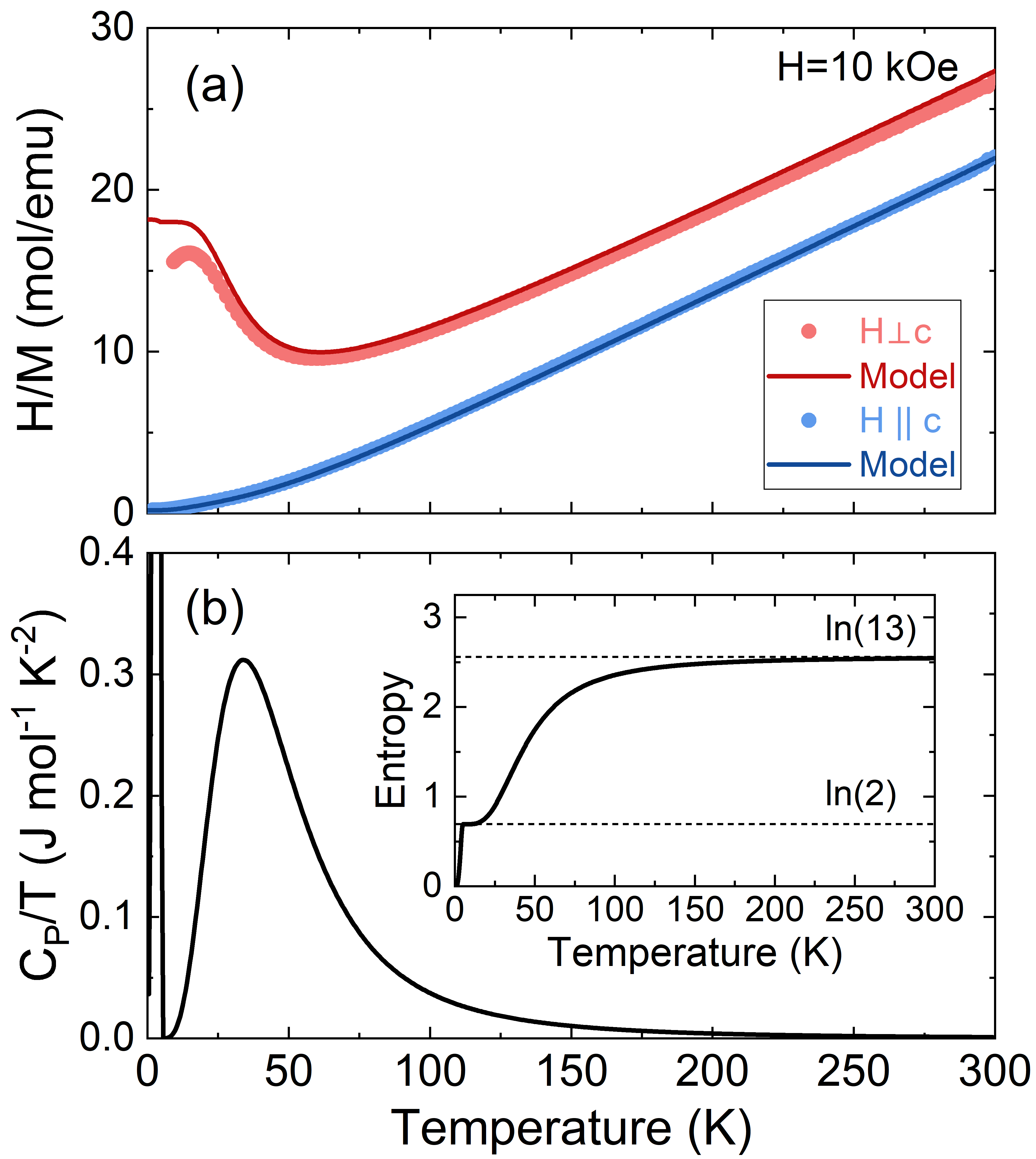}
    \caption{ (a) Transverse ($H\perp c$) and longitudinal-field ($H\parallel c$) magnetization data measured at 10 kOe (circles, plotted as the inverse susceptibility $H/M$) versus temperature.  Solid lines show predictions of the measurement from MF calculations.  (b) MF calculations of the magnetic heat capacity versus temperature.  Inset shows the  magnetic entropy. }
    \label{fig:thermo}
\end{figure}

\subsection{Analysis of raw pulsed field data}

Pulsed field data is proportional to the sample susceptibility, $dM/dH$. A peak in the raw $dM/dH$ data indicates a phase transformation and the peak values are are fit to obtain the critical field, as shown in Fig.~\ref{fig:dMdH}. Magnetization [$M(H)$] data are obtained by subtracting the empty coil signal from the data with sample inside the coil and integrating with respect to field (time).

\begin{figure}      
	\includegraphics[width=0.9\linewidth]{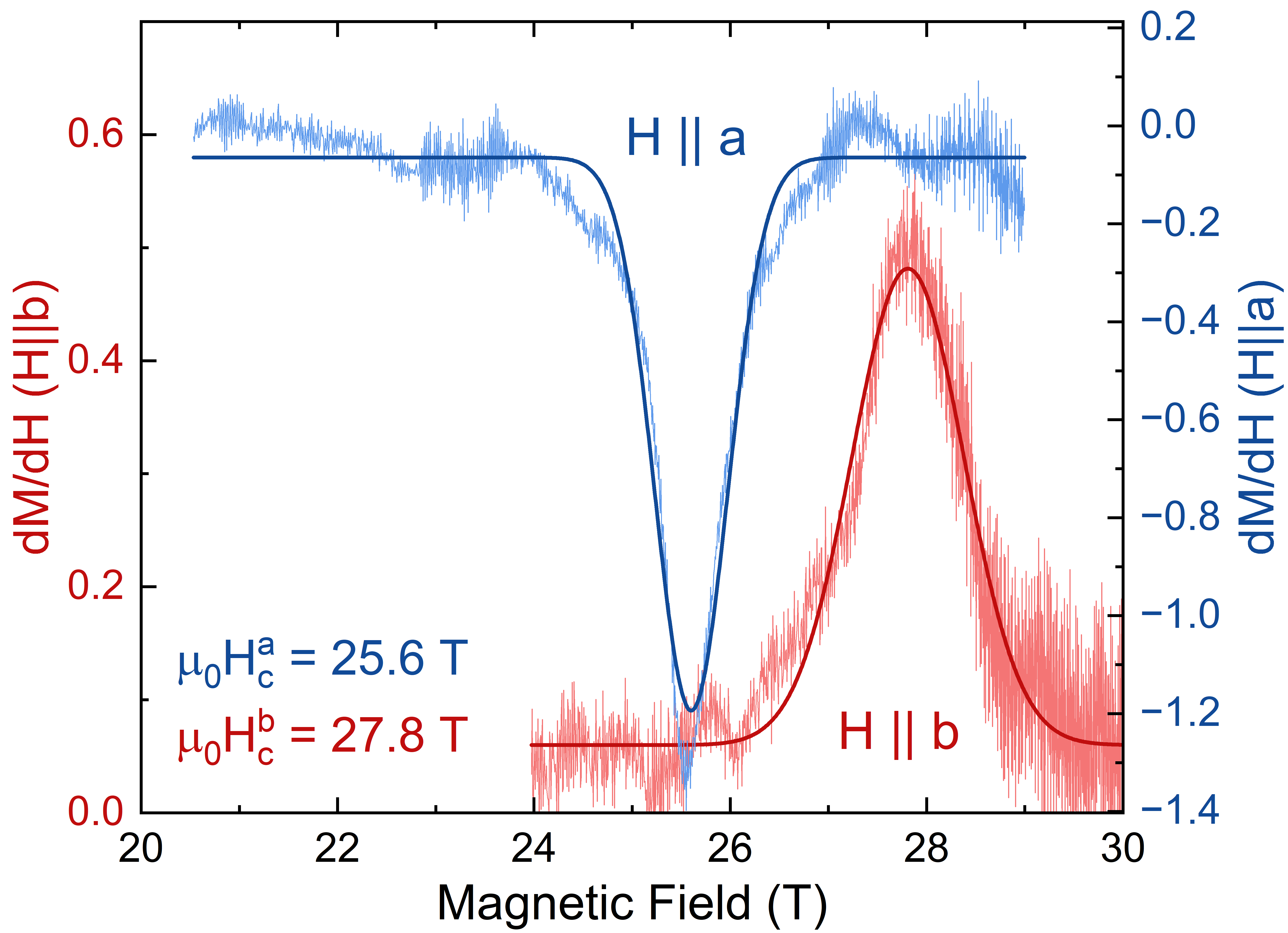}
	\caption{ Raw pulsed-field data proportional to $dM/dH$ in the critical field region. Fits identify the critical field in the $(100)$ and $(210)$ directions.}
 \label{fig:dMdH}
\end{figure}

\subsection{Field rotation and PDO data}
\begin{figure}      
	\includegraphics[width=1.\linewidth]{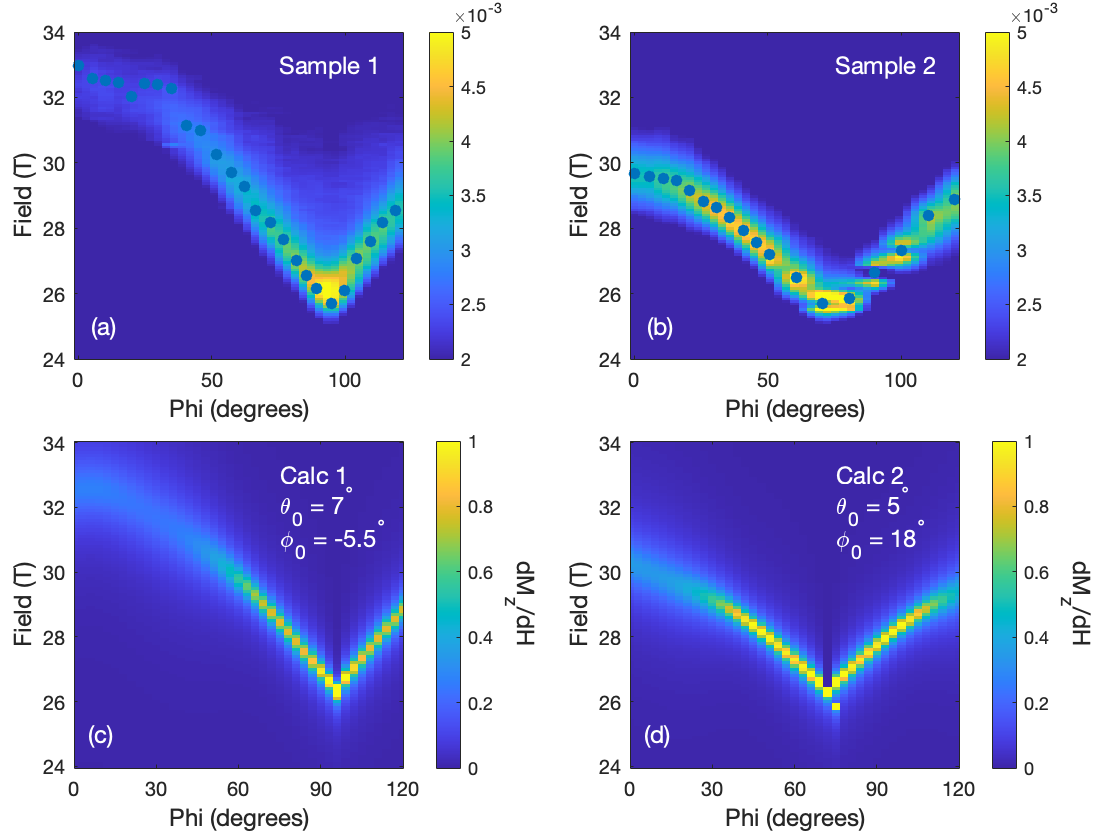}
	\caption{\label{fig:PDO} (a), (b) PDO data taken on two different samples with different tilt angles.  Symbols are fits to the data at different $\phi$ angles. (c), (d) MF calculations of $dM_z/dH$ with angle misalignments, $\theta_0$ and $\phi_0$, that best match the observed data.}
\end{figure}

The proximity detector oscillator (PDO) measurements were performed at the Los Alamos pulsed field facility with fields up to 60 T at 0.6 K. The crystal was placed on top of a 10-turn coil made of 46-gauge copper wire, which is connected to the PDO circuit with resonant frequency in the range of 21 - 36 MHz. The technique is sensitive to changes of electrical conductivity and magnetic susceptibility: $f \propto dM/dH$ \cite{Altarawneh2009, Ghannadzadeh2011}. The rf coil is mounted on goniometer enabling rotation about the coil axis at cryogenic temperatures. This configuration allows the in-plane orientation of applied field to be varied whilst measuring the transverse ($c$-axis component of) magnetic susceptibility.

Fig.~\ref{fig:PDO}(a) and (b) show PDO data for two different sample mountings as a function of planar rotation angle ($\phi$) and field.  Changes in the resonant frequency are associated with changes in the sample susceptibility, $\chi_z = dM_z/dH$ such as that which occurs at the critical spin-reorientation field. For an ideally aligned sample, rotation would occur around the $c$-axis and PDO measurements would adopt a six-fold pattern of the critical field caused by the planar anisotropy parameter $B_6^6$.  However, small tilts of the sample, which can be imposed by large sample torque, will introduce a field component parallel to $c$ that increases the apparent critical field.  Such misalignment will lead to the observed two-fold pattern where the minimum critical field occurs for $\phi$ angles where the field does lie in the $ab$ plane.

We can simulate the effect of sample tilt on the critical field using our magnetic Hamiltonian.  Figures~\ref{fig:PDO}(c) and (d) show MF calculations of $\chi_z = dM_z/dH$ for a misaligned sample as a function of rotation around an axis that is tilted from the crystallographic $c$ axis by $\theta_0$. We introduce misalignment angles to mirror the observed data.  The good agreement indicates that our samples are indeed misaligned to such a degree that we cannot determine $B_6^6$ using this method. However, PDO does determine the minimum critical field that occurs at the one $\phi$ angle where the field lies in the $ab$ plane.  This field (25.7 T) compares favorably to magnetization data shown in the main manuscript.  

\subsection{Sample tilt and magnetization}

The ``transverse" magnetization measurements presented in the body of the paper are magnetizations measured in the direction of the applied field. With perfect sample alignment, one may neglect transverse demagnetization field effects in thin plate-like samples of TbV$_6$Sn$_6$ ($N^{\perp} \approx 0$). However, the sample may experience strong magnetic torques and imperfect alignment, causing the  applied field to have a small longitudinal component. In this case, domain structure and demagnetization fields must be accounted for.

Consider a tilted sample with $\boldsymbol{n}=(\sin \theta_0 \cos \phi_0,\sin \theta_0 \sin \phi_0,\cos \theta_0)$ perpendicular to the $ab$-plane.
The applied field in the sample frame is,

\begin{align}
H_x &= H[\cos \theta_0 +\sin^2 \phi_0 (1-\cos \theta_0)] \\ \nonumber
H_y &= -H\sin \phi_0 \cos \phi_0 (1-\cos \theta_0) \\ \nonumber
H_z &= H\cos \phi_0 \sin \theta_0
\end{align}
If faceting on the sample surface allows for good alignment in the planes, we may assume either $\phi_0=\pi/2$ or $\phi_0=0$, giving sample frame fields of $\boldsymbol{H} = H(1,0,0)$ or $\boldsymbol{H} = H(\cos \theta_0 ,0, \sin \theta_0)$, respectively.  We assume $\phi_0=0$, giving the ``transverse" field a longitudinal component.

Domain structure forms to minimize the magnetostatic energy of a sample. In Ising magnets with highly mobile domain walls, the average internal field is empirically $\overline{H}_{in}^z=H_a^z-N^z \overline{M}^z \approx 0$ \cite{Cooke,Mennenga}, where $\overline{M}^z$ is the average magnetization of the sample and $N^z$ is its demagnetization factor. The experimental susceptibility of a multidomain sample is then $\chi_{exp}^{zz} = \overline{M}^z/H_a^z=1/N^z$. The finite energy of the domain walls, and pinning potentials, will modify this result. In a plate-like sample with $N^z \approx 1$ and highly mobile domain walls with negligible wall energies, we expect $H_a^z \approx \overline{M}^z$ up to the saturation value of the magnetization, $M^z(T, H_x)$. In TbV$_6$Sn$_6$, 
$\mu_0 \boldsymbol{M} = 0.0725 \braket{\boldsymbol{J}}$ Tesla. When the spins are fully polarized ($\braket{J^z} \approx 6$ at $T=0$ K and $\mu_0H_x=0$ T), this gives a saturation field of $\mu_0H_z = 0.4$ T. 

In Fig.~\ref{fig:MagFig}(a) we show the MF magnetization using the Case I CEF parameters (small $B_6^6$) with $\mathcal{J}_0=0.122$ K at $T=0.625$ K. The applied field is along the hard axis assuming a $\theta_0 = 2^{\circ}$ sample tilt. The solid lines show the transverse magnetization and the magnetization in the direction of the applied field, which have a low-field slope of $0.1 \mu_B/{\rm T}$. If we account for domains, we should take the longitudinal component of the magnetization to be the minimum of $\braket{J^z}_{MF}$ and $\overline{\langle J^z \rangle} =\mu_0 \overline{M}^z/0.0725 = B^z/0.0725$, which gives the dashed line.  Its low field slope is $0.12 \mu_B/{\rm T}$, in better agreement with the experimental value. Increasing the sample tilt leads to a steeper rise in the low-field magnetization that is less consistent.

\begin{figure}[htp]
\centering
\begin{minipage}[b]{0.47\linewidth}
\includegraphics[trim={0.4cm 0.1cm 1.1cm 0.8cm},clip,width=4cm]{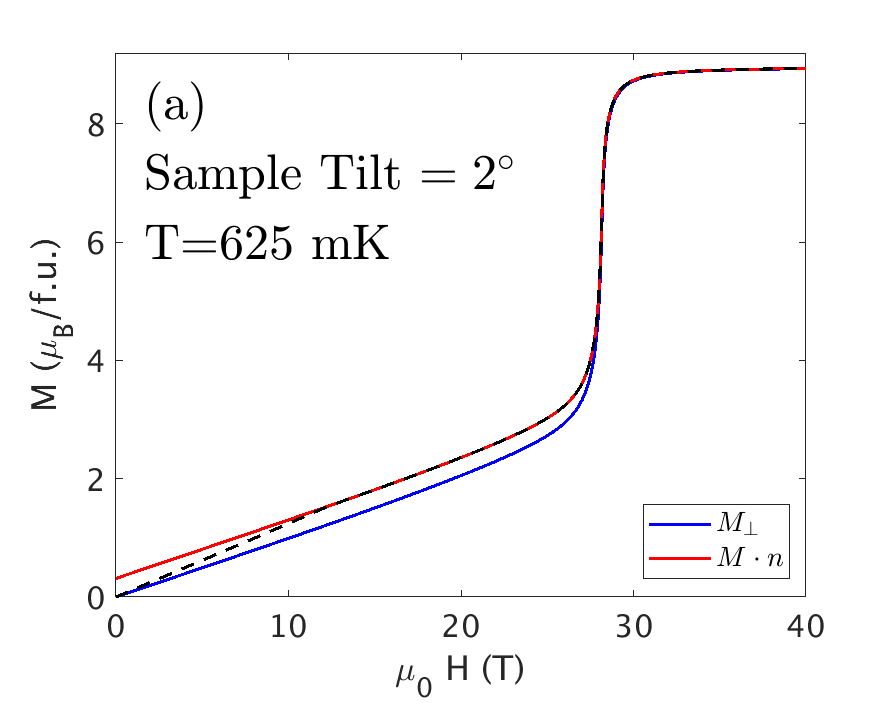}
\end{minipage}
\quad
\begin{minipage}[b]{0.47\linewidth}
\includegraphics[trim={0.4cm 0.1cm 1.1cm 0.8cm},clip,width=4cm]{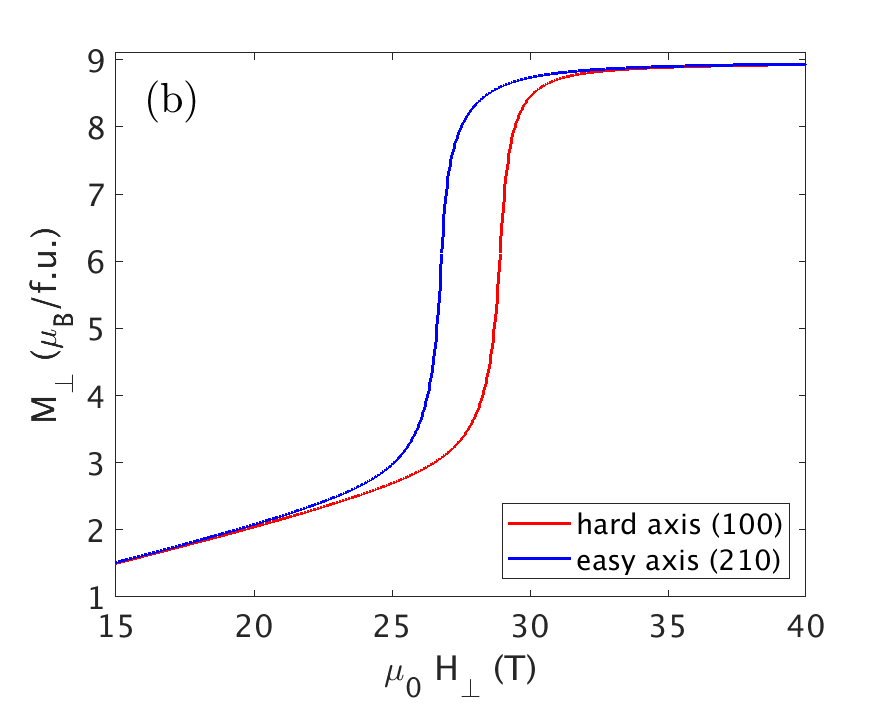}
\end{minipage}
\caption{Effect of sample tilt and domain structure on the MF magnetization calculated for the Case I CEF parameters at $T=625$ mK. The interaction is $\mathcal{J}_0=122$ mK which gives the experimental $T_C \approx 4.4$ K. (a) The MF magnetization along the transverse direction (100), $M_\perp$, and in the direction of the tilted applied field, $\boldsymbol{M}\cdot \boldsymbol{n}$, with a $2^{\circ}$ sample tilt. The dashed line shows the magnetization when $M^z$ is the minimum of either its MF value or $\overline{M^z}$, the average longitudinal magnetization accounting for domain structure. (b) Easy (210) and hard (100) transverse magnetization curves assuming a $2.5^{\circ}$ tilt between measurements. The easy (hard) axes calculations assume an out of plane tilt of  $0.5^{\circ}$ ($3^{\circ}$) in order to roughly mimic Fig. 2(a) in the main text.} 
\label{fig:MagFig}
\end{figure}

The calculated curves, with an identical sample tilt along both the easy and hard axes are inconsistent with the measured magnetizations. In order to obtain the $\approx 2.2$ T separation of the magnetization curves seen experimentally, we must either increase the value of $B_6^6$ or assume the sample tilt changes between the easy and hard axis measurements. In Fig.~\ref{fig:MagFig}(b), we use the Case I CEF parameters to calculate the magnetizations assuming a relative $2.5^{\circ}$ tilt between measurements, which reproduces the experimental magnetization curves shown in Fig.~2(a) of the main text reasonably well; however, if one considers the magnetization in the direction of the applied field, rather than just $M_\perp$, and accounts for domain structure, the agreement is not as good. In particular, we expect the slope of the low-field magnetization to vary with the sample tilt; this is not seen in the experimental data, but it is possible that the sample only tilts or twists out of alignment at high fields.

Similar magnetization curves to those shown in Fig.~\ref{fig:MagFig} can be produced if $B_6^6 = 5\times 10^{-6}$ meV, above the QTCP threshold, but there are substantial discrepancies if $B_6^6$ is increased above this value. With the field along the easy axis, the rise in the magnetization near the phase transition is less steep than what is observed experimentally. Along the hard axis, the magnetization rises more steeply, cutting through the easy axis curve (as in Fig.~2(f) of the main text). It takes a substantial sample tilt to eliminate this crossing, and the separation of the two curves is larger than what is measured experimentally.

\subsection{Estimating CEF parameters from magnetization data}
We can use the magnetization data to constrain our estimates of the CEF parameters. We start with the classical magnetic anisotropy energy (MAE) 
\begin{equation}
    E_A = K_1\sin^2\theta + K_2\sin^4\theta+K_3\sin^6\theta+K_3'\sin^6\theta\cos6\phi
\end{equation}
where $\theta$ and $\phi$ are the spherical angles describing the Tb moment direction. The MAE constants are related to the CEF parameters according to the following relationships
\begin{align}    
\label{eqn:MAE_const}      
&K_1=-3J^{(2)}B_2^0-40J^{(4)}B_4^0-168J^{(6)}B_6^0 \\ \nonumber
&K_2=35J^{(4)}B_4^0+378J^{(6)}B_6^0 \\ \nonumber
&K_3=-231J^{(6)}B_6^0 \\ \nonumber
&K_3'=J^{(6)}B_6^6 \\ \nonumber       
\end{align}
where $J^{(2)}=J(J-\frac{1}{2})$, $J^{(4)}=J^{(2)}(J-1)(J-\frac{3}{2})$, and  $J^{(6)}=J^{(4)}(J-2)(J-\frac{5}{2})$.  In a transverse field along $x$, the Zeeman energy is $E_Z=-\mu_0MH\sin\theta$ where $M=9 ~\mu_B$ is the full Tb moment. We can determine the equilibrium condition for the Tb angle by minimizing the total energy $E=E_A+E_Z$, leading to 
\begin{equation}
    2K_1\sin\theta + 4K_2\sin^3\theta+6(K_3-|K_3'|)\sin^5\theta - \mu_0MH=0
    \label{eqn:MAE_deriv}
\end{equation}

For a small transverse field, the moment will tilt towards the field direction and the equilibrium condition for a small tilting angle is $\theta\approx \mu_0MH/2K_1$.  We can calculate the effective low-field transverse susceptibility in this limit as $\chi_x=M_x/\mu_0 H = M^2/2K_1$.  Thus, $\chi_x$ can be used to determine $K_1$. From fits to our DC susceptibility measurements, we estimate that $\chi_x=0.12~\mu_B/T$ which gives $K_1 = 20$ meV. Note that for a uniaxial magnet in the absence of a molecular field, the main INS $|6\rangle\rightarrow|5\rangle$ transition out of the ground state has an energy of $\Delta \approx 2K_1/J = 7.9$ meV and from this we estimate that $K_1 = 23.7$ meV.

We can also study the equilibrium condition that leads to the magnetization jump at a critical field $H_c$ where the moment lies fully in the plane.  At $H_c$, the energy at the critical angle $\theta_c$ is equal to the energy of the in-plane state with $\theta=\pi/2$. 
\begin{align}
\label{eqn:critical}
    & K_1\sin^2\theta_c + K_2\sin^4\theta_c + (K_3-|K_3'|)\sin^6\theta_c \\ \nonumber
    & -\mu_0MH_c\sin\theta_c - (K_1+K_2+K_3-|K_3'|-MH_c) = 0\\ \nonumber
\end{align}
Here, we can use the experimental values of $K_1$ and $H_c$ as constraints to numerically solve Eqns.~\ref{eqn:MAE_deriv} and \ref{eqn:critical} for $\theta_c$, $B_4^0$, and $B_6^0$ while letting $B_6^6=0$. To do this, we also need another equation.  We can use the relation for the $|5\rangle \rightarrow |4\rangle$ energy splitting $\Delta'=4.9$ meV.
\begin{align}
       \Delta' &= -3B_2^0(2J-3)-40B_4^0(2J-3)(J-1)(J-5) \\ \nonumber
       &-42B_6^0(2J-5)(2J-3)(J-2)(J-1)(2J-21) \\ \nonumber
\end{align}
Using the three observables ($K_1=23.7$ meV, $\mu_0H_c=25.6$ T, and $\Delta'=4.9$ meV), we simultaneously solve the three equations to obtain; $\sin\theta_c = 0.297$, $B_4^0=-6.16\times 10^{-4}$ meV, and $B_6^0 = 2.43\times 10^{-6}$ meV.  We also use the $K_1$ constraint to obtain $B_2^0 = -0.0975$ meV.  These values are very similar to the CEF parameters obtained from fitting the INS data directly. 

\section{Extended modeling}

\subsection{Hyperfine interactions}

 Terbium has a single stable isotope with nuclear spin $I=3/2$. The strength of the hyperfine interaction, $H_{hyp} = A \boldsymbol{I} \cdot \boldsymbol{J}$, is $A=25.4$ mK \cite{Bleaney}. For comparison, the strongest rare-earth hyperfine interaction is in holmium with $A=39$ mK and $I=7/2$, which leads to substantial corrections to the low-temperature phase diagram of LiHoF$_4$ \cite{Bitko}. The strong Ising anisotropy of LiHoF$_4$ through the critical transverse field leads to a strongly anisotropic effective hyperfine interaction \cite{Chakraborty, Schechter2, McKenzie}. In TbV$_6$Sn$_6$, the anisotropy of the system evolves with the transverse field, and with the INS-derived CEF parameters, the system loses its strong Ising character around 25 T, just below the critical transverse field, which means the hyperfine corrections to the TbV$_6$Sn$_6$ phase diagram are small, with the increase of the critical field at $T=0$ K being $\sim 0.1\%$.

\subsection{Dipolar interactions}

The dipole-dipole interactions between Tb moments are
\begin{align}
\mathcal{H}_{dip} = -\frac{\mathcal{J}_{D}}{2} \sum_{i \neq j} \sum_{\mu \nu} 
D_{ij}^{\mu \nu} J_i^{\mu} J_j^{\nu},
\end{align}
where
\begin{align}
D_{ij}^{\mu \nu} = \frac{1}{r_{ij}^3}
\biggr(\frac{3r_{ij}^{\mu}r_{ij}^{\nu}}{r_{ij}^2}-\delta_{\mu \nu}\biggr)
\end{align}
and $\mathcal{J}_D = \mu_0 (g_J\mu_B)^2/(4\pi)$. Note that at zero wave-vector the off-diagonal components of the dipolar sums vanish due to the lattice symmetry so that these terms may be neglected in mean-field calculations. The Tb are situated on triangular lattices in the basal plane, with the lattice spacing at $T=10$~K being $a=b=5.512~\text{\AA}$ and $c=9.166~\text{\AA}$ \cite{PhysRevMaterials.6.104202}, leading to a spin density, $\rho_s=4.146 \times 10^{27}~\text{Tb/m}^3$. Dipolar interactions on this lattice favor antiferromagnetic alignments, with a characteristic strength, $\rho_s \mathcal{J}_D=5.8$ mK, while the spins experimentally order ferromagnetically at $T_{\rm C} \approx 4.4$ K \cite{PhysRevMaterials.6.104202}, indicating that the dominant coupling between spins is not dipolar, but instead indirect exchange mediated by the vanadium conduction electrons. 

\subsection{Doublet splitting in transverse-field}

The primary effect of $B_6^6$ is to increase the splitting of the effective ground state doublet along the easy (210) direction.  The behavior along the hard (100) direction is more complicated, as the splitting first decreases as a function of $B_6^6$ before effectively passing through zero and growing again, leading to the second critical $B_6^6$ where the hard axis transition also becomes second order.
Fig.\ref{fig:delta166} shows the behavior of this splitting as a function of the easy (210) axis field and $B_6^6$, where we use the invariance of the crystal field Hamiltonian under a rotation of $2\pi/6$ combined with a change of the sign of $B_6^6$ to capture the hard (100) axis behavior using negative $B_6^6$.  The white region indicates where the splitting passes through zero.

\begin{figure}[htp]
\centering
\includegraphics[width=9cm]{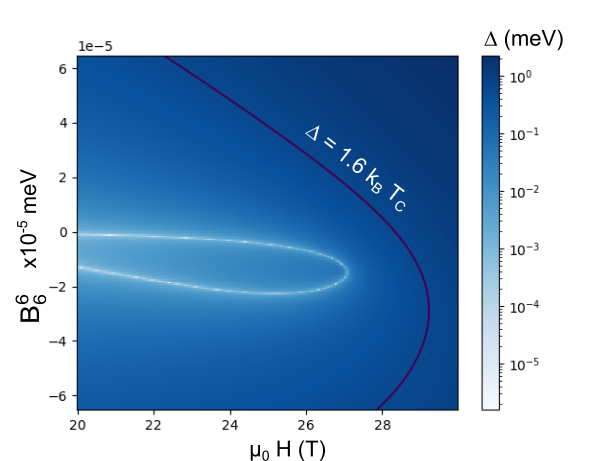}
\caption{Effective splitting of the ground state doublet as a function of $B_6^6$ and $\mu_0 H$.  For positive signs of $B_6^6$, the field is along the (210) easy axis; swapping the sign of $B_6^6$ rotates the system by $2\pi/6$ and gives the behavior along the (100) hard axis  instead.  The black line indicates the splitting expected to correspond to the second order critical field, $\Delta = 1.6 k_B T_c$, if it is not preempted by a first-order metamagnetic transition.  The white line indicates where the splitting reaches zero.}
\label{fig:delta166}
\end{figure}

\section{Density-functional theory calculations}
Given the challenges in experimentally determining $B_6^6$, we conducted density-functional theory (DFT) calculations to estimate this parameter.
The ab initio estimation of CEF parameters is generally challenging in MF methods like DFT, as the $4f$ self-interaction in DFT is at least an order of magnitude larger than the crystal field strength itself.
In this study, we employed a recently proposed total-energy-mapping method that leverages the invariance of the orbital dependence of self-interaction errors~\cite{lee2024arxiv}.

For the hexagonal system with $C_{3v}$ point-group symmetry at the $R$ site, the in-plane anisotropy within CEF theory depends solely on $B_6^6$.
\begin{equation}
E(\phi,\theta=\frac{\pi}{2}) = \frac{1}{16}{B_6^6 \langle \mathcal{O}_6^6 \rangle} \cos6\phi
\end{equation}
Here, the nearest $R$-V bond direction in the basal plane is along the [210] direction (defined as $\phi=0$), while the nearest Tb-Tb is along the [100] direction ($\phi=-\pi/6$).
We calculated the total energy difference between these two directions, $\Delta E = E_{210} - E_{100}$, to extract the anisotropy and determine $B_6^6$.

Self-consistent DFT+$U$ calculations, including spin-orbit coupling, were performed.
A sufficiently large $U$ value is necessary to push the $4f$ states deeper away from the Fermi level.
This ensures that the calculations converge to the Hund's rule ground state of Tb$^{3+}$, consistent with experimental results, avoiding the unphysical DFT+$U$ ground state of $|m_l^\downarrow=2\rangle$ caused by
the orbital dependence of self-interaction errors~\cite{lee2024arxiv}.
This approach has been shown to accurately reproduce results for all existing RMn$_6$Sn$_6$ compounds and other well-studied rare-earth-containing isostructural materials.
Further details of the DFT+$U$ calculations are available in other sources.

The calculated energy difference at $U$=10 eV is $\Delta E=0.408$ meV, yielding a $B_6^6$ of $2.08 \times 10^{-5}$meV, which aligns closely with the $B_6^6$ value of the Case II (larger) experimental set.
We also explored the sensitivity of $B_6^6$
to the U parameter. Across a reasonable range of
U values of 6--12 eV, $B_6^6$ varies from
$1.67\times10^{-5}$ to $2.77\times10^{-5}$ meV.
These DFT+$U$ calculations are consistent with the experimental data, supporting the presence of a sizable $B_6^6$.

\bibliography{TbV6Sn6Bib}

\clearpage

\raggedright